\documentclass[twocolumn,prb,aps,amsmath,amssymb]{revtex4-1}
\usepackage{amsmath,amssymb,bm,epsfig}
\usepackage{dcolumn}% Align table columns on decimal point
\usepackage{bm}% bold math
\usepackage{feynmf}
\begin{document}

\title{Infrared Dynamics of  Cold Atoms on Hot Graphene Membranes}

\author{Sanghita Sengupta, Valeri N. Kotov,  and Dennis P. Clougherty}

\affiliation{
Department of Physics,
University of Vermont, 
Burlington, VT 05405-0125}

\date{\today}

\begin{abstract}
 We study the infrared dynamics of low-energy atoms interacting with a sample of suspended graphene at finite temperature.
The dynamics exhibits severe infrared divergences order by order in perturbation theory as a result of the singular nature
of low-energy flexural phonon emission. Our model can be viewed as a two-channel generalization of the independent boson model
with asymmetric atom-phonon coupling. This allows us to take
advantage of the exact non-perturbative solution of the independent boson model in  the stronger channel while
treating the weaker one perturbatively. In the low-energy limit, the  exact solution can be viewed as a resummation  (exponentiation) of the most
divergent diagrams in the perturbative expansion. As a result of this procedure, we obtain
the atom's Green function which we use to calculate the atom damping rate, a quantity equal to the  quantum sticking
 rate.  A characteristic feature of our results is that the Green's function retains a weak, infrared cutoff dependence that
reflects the reduced dimensionality of the problem.  As a consequence, we predict a measurable dependence of the sticking rate on graphene sample size.  We provide detailed predictions for the sticking rate  of atomic hydrogen as a function of temperature and sample size. 
The resummation yields an enhanced sticking rate relative to the conventional
Fermi golden rule result (equivalent to the one-loop atom self-energy), as higher-order processes increase damping at finite temperature.

\end{abstract}

\maketitle
\section{Introduction}

%============ general introduction

The study of the dynamics of cold atoms near suspended graphene samples presents opportunities to explore some of the foundational concepts of quantum mechanics.   Since the binding energy of an atom on graphene is comparable to graphene's quantum excitation energies, the dynamics depends on a quantum treatment of excitations.  Even the atomic motion must be treated quantum mechanically for sufficiently cold atoms.   Thus the theory of cold atom adsorption on graphene must be a fully quantum theory.  

%We also note that mesoscopic graphene samples offer the possibility that finite-size effects appear in the dynamics.

From a practical perspective, information gleaned from dynamical studies of cold atoms near surfaces will likely find use in
the development and refinement of  systems and devices for quantum sensing and information processing.  
In recent years, there have been many experimental advances in the cooling and control of atoms
and molecules.
Picokelvin sources of helium
atoms can now be experimentally prepared \cite{ultracold}, and new technologies
propose to use quantum states of cold atoms and molecules to store and process
information \cite{chuang, lukin}. One example of these potential  applications  is realized in the ``atom chip,'' a microelectronic device
where currents flowing through nanowires generate magnetic fields to process
information stored in the quantum states of cold atoms.  

A second example is in quantum metrology. This includes applications such as recently developed chip-scale atomic clocks  using
cold $^{87}$Rb atoms \cite{atomic-clock}. The operation of these devices will
be strongly impacted \cite{hrs} by how cold atoms and molecules interact with surfaces.  Hence our theoretical studies will impact performance and design of these emerging applications; for example, unintended adsorption of alkali metal atoms leads to the so-called ``patch
effect,'' where random islands of dipoles can cause a rapid dephasing of
entangled states of atoms trapped above  surfaces.

Another potential application is in ``atom optics'' where matter-waves play the
role conventionally performed by laser light in optical systems \cite{interferometer}.  Enhancing the
reflection of matter waves from surfaces might be used to make low-loss atomic
mirrors  \cite{atom-mirror}, waveguides for atom interferometers
\cite{interferometer} or microtraps for the quantum information processing of
cold atoms \cite{microtrap}.

Research in the field of cold atom-surface interactions traces back to the beginning years of quantum theory, where the threshold behavior for quantum adsorption was first explored by Lennard-Jones \cite{lj3}. Early theoretical work concluded that the sticking probability s(E) of a particle with incident energy E near threshold is directly proportional to the square of the transition matrix element and varies inversely with the incident particle flux, such that s(E) $\propto$ $\sqrt {E}$. More recent studies \cite{dpc92, dpc10,dpc11} have predicted new scaling laws for neutral and charged particles based on quantum many-body effects, such as   orthogonality catastrophe-type phenomena.

%---added by Valeri from proposal, to stimulate longer general intro

For inelastic interactions, the primary channel of energy exchange involves the creation and annihilation of phonons.  Free-standing graphene has two in-plane acoustic modes and one out-of-plane flexural mode.
The flexural mode has a quadratic dispersion  near
the zone center; however, under uniform tension, the
flexural dispersion becomes linear, leading to vanishing phonon density of states (DOS) for suspended graphene. 
%  This is the case for suspended graphene.  Thus, the
%phonon density of states (DOS)  for suspended graphene vanishes
%linearly with frequency near zero frequency.  
  In contrast to a constant DOS at zero frequency, a linear DOS eases a
well-known divergence in the displacement autocorrelation function of the
$n^{\mathrm{th}}$ nearest neighbors $\langle (u_n-u_0)^2\rangle$ in two dimensions  \cite{mermin}
and stabilizes the suspended layer mechanically, circumventing the
``crumpling'' instability.
In the case of inelastic atom-graphene interactions, vestiges of this divergence can be found in a perturbative expansion of the atom's self-energy \cite{dpc12}. The linear DOS of the flexural phonons, when combined with the frequency-dependent atom-phonon
coupling gives a (log) divergent atom self-energy  at zero temperature.  This implies that without a low-frequency cut-off, the second-order shift in
the binding energy  of a light atom on suspended graphene is formally
divergent \cite{dpc12}.  Recent numerical calculations on the physisorption of atomic hydrogen to suspended graphene \cite{lepetit-jackson} do not take into account the effect of this infrared
divergence which remains a  theoretical challenge.

%New techniques that lead to the production of cold molecules may also usher in
%a new regime for doing chemistry \cite{coldchem} where reactions are selected
%by controlling the quantum state of the reactants at one-millionth of a degree
%above absolute zero.  Ultracold surface chemistry and catalysis offer new
%quantum phases of matter that could be used for qubits in a quantum computer
%\cite{qubit}  or for precision tests of the standard model \cite{hinds}.
%Other potential applications include controlled doping for novel nanoelectronic
%devices \cite{me}, precision mass sensors \cite{sensor}, switches, mechanical
%resonators \cite{mceuen} and other nanomechanical devices \cite{singh}. 

%=========== this work

In the present work, we use a diagrammatic approach to study the dynamics of  quantum sticking near graphene membranes.
The quantum sticking probability is related to the damping rate of the atom, calculated from its self-energy. 
 In fact, the study of atom self-energy to probe various aspects of atom-surface scattering is not a recent one and has been explored before \cite{annett,koonin,light95}. The presence of  infrared divergent terms, due to many emitted low-energy flexural phonons, 
 has been previously pointed out \cite{dpc12,dpc13},  
 and our goal in this work is to provide a systematic study of these effects in the context of renormalized perturbation theory.

We will consider membranes at  finite temperature T,  where  the infrared divergence  problem is especially severe since the number of thermally-activated
phonons tends to infinity with increasing membrane size. However, there is a well-defined way to take into account (through resummation) such infrared-singular  processes; especially suggestive is an exact solution for a particle (atom) interacting
with a bath of phonons (independent boson model  \cite{mahan}), even though our problem is not exactly solvable in the same sense due to the presence
of two types of atom-phonon couplings \cite{dpc13}. 

Quite remarkably, the problem under consideration is also technically  similar to the infrared problems 
present in finite-T  ``hot" quantum electrodynamics (QED) and quantum chromodynamics (QCD) 
 due to the long-range, unscreened, nature of gauge interactions \cite{qed,qed1,qed2,qed3}.
 These problems are usually resolved within the  finite T generalization of the so-called 
 Bloch-Nordsieck scheme  \cite{qed,qed1}, which extracts the exact
 infrared behavior of the theory by summing the most important Feynman diagrams; it is
  quite similar  to the exact solution of the   independent boson model (IBM)  for single atom -- phonon bath 
 mentioned previously. Physically this corresponds to the correct account of the damping
 provided by many emitted phonons, and we will implement similar schemes to obtain the
 sticking rate of adatoms (equivalent to calculating the atomic damping rate).
 
 Since the infrared divergences are caused by emitted low-momentum phonons $q \rightarrow 0$, at any finite-T around the Debye temperature, the system is effectively in the high temperature (hot) limit as far as its infrared properties
 are concerned, $T \gg v_s q$, where $v_s$ is the flexural phonon speed. 
Similar to the case of hot QED,   this allows for a particularly theoretically clean and elegant way to perform resummation 
 of the leading infrared divergences which come as powers of logarithms. 

We present a detailed study of the atomic self-energy at one and two loops, and subsequently perform
resummation of the leading infrared terms in the spirit of the IBM or hot QED,  but adapted for our two-channel situation.
This allows us to obtain reliable non-perturbative results for the damping rate as a function of temperature (T) as well as membrane size (L),
which serves as the effective infrared cutoff in the problem. The description of these dependencies and the theoretical methodology
we use to calculate them is the main goal of this work.

The rest of the paper is organized as follows.
In Section~\ref{sec:model}, we introduce the effective model of atoms interacting with graphene membranes. 
In Section~\ref{sec:div}, we analyze the infrared divergences that appear up to two-loops and then present
the non-perturbative solution in the spirit of  the IBM  in 
Section~\ref{sec:ibm}. 
Our results for a H atom's sticking rate as a function of temperature and size are collected in   Section~\ref{sec:stick}. 
Section~\ref{sec:conclusions} contains our conclusions. Some technical aspects of high-order perturbation theory are
presented in Appendices A and B.
%==========================

\section{Model of cold atoms on graphene membrane at finite temperature}
\label{sec:model}

\begin{figure}[h]
\center
\includegraphics[width=0.8\columnwidth]{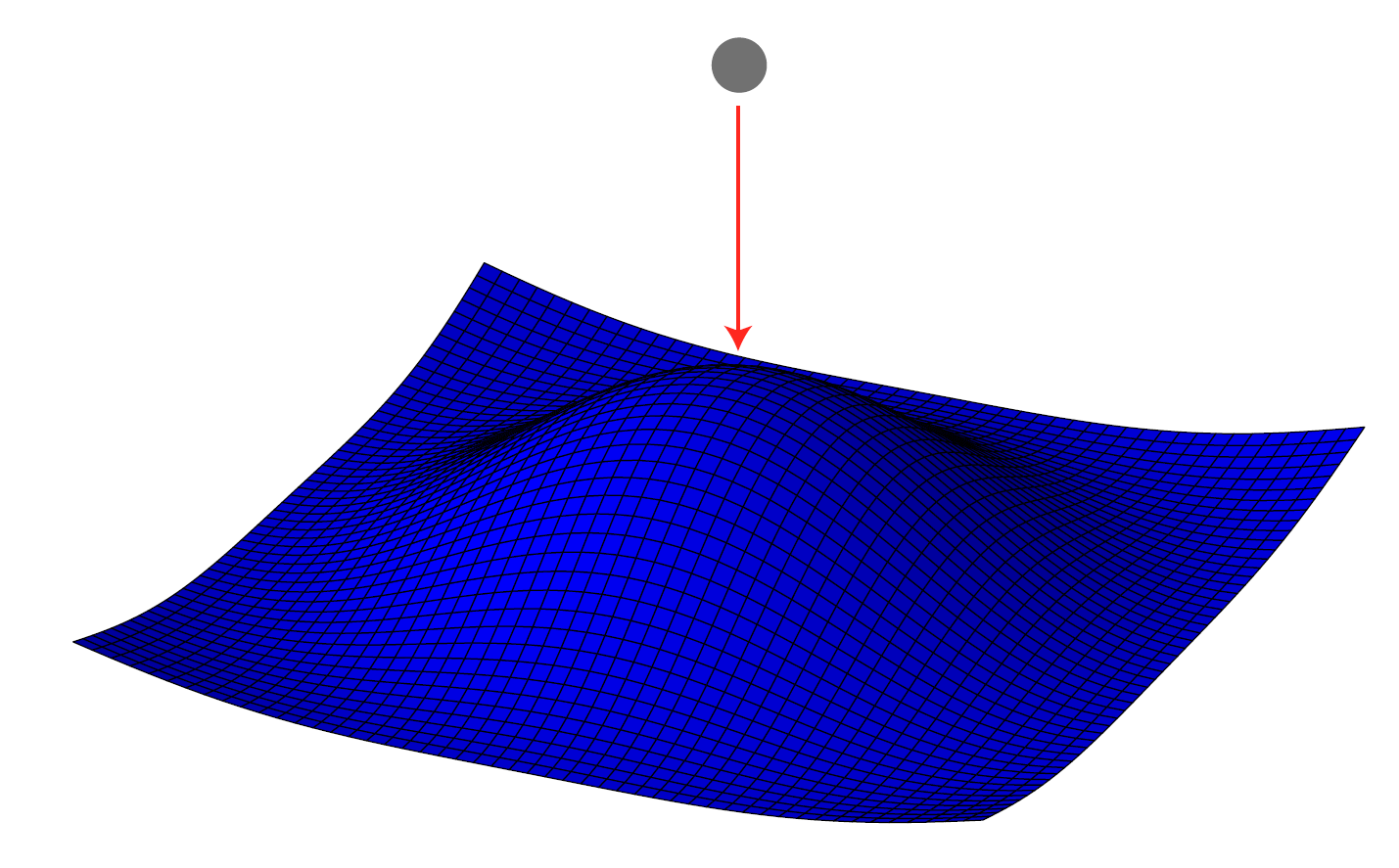}
\caption[fig]{Sketch of the membrane with an impinging atom. The membrane distorts from its initial equilibrium plane with the presence of the adatom.}
\label{fig:model}
\end{figure}

We study the interaction of a cold atom with a clamped, elastic membrane (Fig.~\ref{fig:model}), where the inelastic interaction between the atom and the membrane occurs through the creation and annihilation of flexural phonons of the membrane. For atoms focussed near the center of the suspended membrane, the circularly symmetric modes dominate the inelastic scattering. Thus we consider the atom interactions with only the axisymmetric ($m=0$) modes \cite{dpc13}.

The Hamiltonian of the system is written as $H = H_p +H_{ph} +H_{c}$. The terms represent the Hamiltonian for the particle, phonon-bath and the coupling term, respectively, and are given as:
\begin{equation}\label{1}
H_{p} = E_{k}c_{k}^{\dagger}c_{k}-E_{b0}b^{\dagger}b 
\end{equation}
\begin{equation}\label{2}
H_{ph} = \sum_{q}\omega_{q}a_{q}^{\dagger}a_{q}
\end{equation}
\begin{equation}\label{3}
H_{c} = -g_{kb}(c_{k}^{\dagger}b + b^{\dagger}c_{k})\sum_{q}\xi_{q}(a_{q}+a_{q}^{\dagger}) - g_{bb}b^{\dagger}b\sum_{q}\xi_{q}(a_{q}+a_{q}^{\dagger})
\end{equation}

Here, $c_{k}$ $(c_{k}^{\dagger})$ annihilates (creates) a particle in the entrance channel $|k\rangle$ with energy $E_{k}$;  $b$ $(b^{\dagger})$ annihilates (creates) a particle in the bound state $|b\rangle$ with energy -$E_{b0}$ in the  potential of a static membrane in its initial equilibrium position; $a_q$   $(a_q^{\dagger})$ annihilates (creates)  a phonon in the membrane with energy $\omega_q$;  $g_{kb}$ is the coupling strength of phonon-assisted transitions of the atom between continuum $|k\rangle$ and bound state $|b\rangle$; $g_{bb}$ is the coupling strength of the bound atom to flexural phonons. The form of $\xi_{q}$ depends on the specific particle-excitation coupling, and particularly in this model, $\xi_{q}$ is $q$-independent\cite{dpc13}.

%\includegraphics[width=0.9\columnwidth]{model}\\

%\noindent
%FIG. 1: Sketch of the circular membrane of radius a with an impinging atom of mass M. The membrane distorts out of the xy plane in the presence of the adatom. Each differential patch of the membrane, located at (r, $\theta$, u) (polar co-ordinates), contributes to the atom-membrane interaction \cite{dpc13}.\\

We take the membrane to be initially in  thermal equilibrium with temperature $T$, while the atoms have an initial energy of $E_k$. The atom self-energy must be calculated using non-equilibrium Green functions (Keldysh or the contour-ordered Green functions). The Feynman rules using the Keldysh formalism are summarized below for our model:
\begin{itemize}
\item The solid dot corresponds to the interaction vertex $g_{kb}$.
\item The open dot corresponds to the interaction vertex $g_{bb}$.
\item Lines labeled by $\emph{b}$ correspond to the bare (retarded) Green function in the bound state $G_{bb}^{r} (E) = 1/(E+E_{b0}+i\eta)$
\item Lines labeled by $\emph{k}$ correspond to the bare (retarded) Green function of the atom in the continuum state $G_{kk}^{r} (E) = 1/(E-E_{k}+i\eta)$. 
\item Wiggly lines correspond to the phonon propagators and are given as: \\
 $D^{<}(\omega) = -2\pi i\sum_q [(N_{q}+1)\delta(\omega+\omega_{q}) + N_{q}\delta(\omega-\omega_{q})]$ and $D^{r}(\omega) = \sum_q [1/(\omega-\omega_{q} + i\delta) - 1/(\omega+\omega_{q} + i\delta)]$. \\
 Here, $N_{q}$ is the equilibrium phonon occupation number $N_q=1/(e^{\beta\omega_{q}}-1)$, and $\omega_{q} = v_s q$ 
 for a membrane under tension.

\item Each diagram is weighted by $(i/\hbar)^n$, where n is the number of phonon loops.
\end{itemize}

Two additional comments are in order.
First, we also assume that, by definition,  the ground state is the initial (symmetric) vacuum of the phonons.  
It is possible that at low temperature, the phonons could condense, leading to finite membrane displacement and a symmetry-broken state;
such a scenario was considered previously by one of us  within a mean-field theory approach applied
to this model\cite{dpc13}. The present study can not shed light on that result,  since 
we aim to collect the leading infrared-divergent terms that happens essentially in the ``high-temperature"
regime ($T \gg v_s q$, and $T > E_b$) and we are under the (well-satisfied) weak-coupling condition
$\frac{g^2}{E_b} \ll 1$ (and  $\frac{g^2T}{E_b^2} \ll 1$), where $g^2$ is either $g_{kb}^2$ or $g_{bb}^2$, with dimension of energy when 
appropriately written (see below).
Under these assumptions,  a perturbative expansion around the symmetric vacuum seems well justified; however, we certainly can not rule out the possibility of symmetry breaking.

Finally, we mention that the diagram technique constructed above is completely equivalent
to simply working with the real-time finite temperature Green's function for the phonons \cite{mahan,landaulifshitz10}.
This can be easily seen by examining the structure of  the one-loop result   Eq.~\eqref{8} and the way
it follows from  Eq.~\eqref{5}. Furthermore, it is clear that for the purposes of extracting the leading, infrared-divergent logarithmic terms, it
is sufficient to use the small momentum/high temperature limit of the phonon propagator  in the form $D^<(\omega, q) = -2\pi i (T/\omega_{q})[\delta(\omega+\omega_{q}) + \delta(\omega-\omega_{q})]$.

%\\

\begin{figure}
\center
\begin{fmffile}{fgraphs}
%(along, up)
    \begin{fmfgraph*}(100,100)
      \fmfleft{i1}
      \fmfright{o1,o2}
      \fmfright{p}
      \fmf{fermion,label=$|k\rangle$}{i1,w1}
      \fmf{fermion,label=$|b\rangle$}{w1,p}
      \fmf{photon,label=$\omega$}{w1,o2}
      \fmfdot{w1}
      \fmfv{lab=$g_{kb}$,lab.dist=-0.15w}{w1}
    \end{fmfgraph*}
    \hspace{2mm}
    \begin{fmfgraph*}(100,100)
    \fmfleft{i1}
      \fmfright{o1,o2}
      \fmfright{p}
      \fmf{fermion,label=$|b\rangle$}{i1,w1}
      \fmf{fermion,label=$|b\rangle$}{w1,p}
      \fmf{photon,label=$\omega$}{w1,o2}
      \fmfblob{0.05w}{w1}
      \fmfv{lab=$g_{bb}$,lab.dist=-0.15w}{w1}
    \end{fmfgraph*}
\end{fmffile}
\caption{Feynman diagrams with two kinds of vertices: transition of atom from $|k\rangle$ to $|b\rangle$ state via a phonon has vertex $g_{kb}$ (left), and atom-phonon interaction in the bound state $|b\rangle$  has vertex $g_{bb}$ (right). }
\label{fig:rules}
\end{figure}
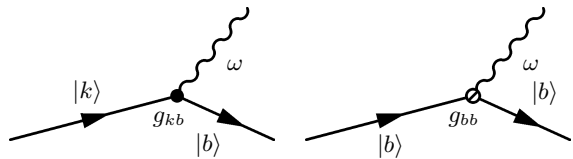

\section{Infrared-divergent self-energies at low orders}
\label{sec:div}

\subsection{1-loop Atom Self-Energy}
Applying the above  Feynman rules, we find the finite temperature atom self-energy at the  1-loop level is given by \\

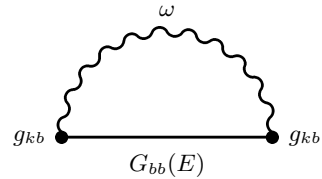
\begin{figure}
\center
\vspace{10mm}
%\hspace{2cm}
\begin{fmffile}{digram}
\begin{fmfgraph*}(80,80)
\fmfleft{i1}
\fmfdot{i1}
\fmfright{i2}
\fmfdot{i2}
\fmf{plain,label=$G_{bb}(E-\omega)$}{i1,i2}
\fmf{photon,left,label=$\omega$}{i1,i2}
\fmflabel{$g_{kb}$}{i1}
\fmflabel{$g_{kb}$}{i2}
\end{fmfgraph*}
\end{fmffile}
\caption{One-loop atom self-energy $\Sigma_{kk}^{(r)}$.}
\label{fig:oneloop}
\end{figure}

\begin{equation}\label{4}
\begin{split}
\Sigma_{kk}^{(r)}(E)&=i\int\frac{\mathrm{d}\omega}{2\pi}\sum_{q}g_{kb}^{2}\xi^{2}[G_{bb}^{<}(E)D^{r}(\omega)\\
&\quad + G_{bb}^{r}(E)D^{<}(\omega) + G_{bb}^{r}(E)D^{r}(\omega)]
\end{split}
\end{equation}
Since we take the incoming particle to be out of equilibrium with the phonon bath with the Green function $G_{bb}^{<}(E) =0$,  we obtain:
\begin{eqnarray}\label{5}
\begin{split}
\Sigma_{kk}^{(r)}(E) &= i\int\frac{\mathrm{d}\omega}{2\pi}\sum_{q}g_{kb}^{2}\xi^{2}[G_{bb}^{r}(E)D^{<}(\omega)\\&\quad + G_{bb}^{r}(E)D^{r}(\omega)]
\end{split}\\
%\end{equation} 
%\begin{equation}\label{6}
\begin{split}
 &= g_{kb}^{2}\xi^{2}\sum_{q}\bigg[\frac{N_{q}}{E+E_{b0}+\omega_{q}-i\eta}\\
& \quad + \frac{N_{q}+1}{E+E_{b0}-\omega_{q}+ i\eta}\bigg]
\end{split}\nonumber
\end{eqnarray}

Next, in the continuum limit, we obtain the real and the imaginary part of $\Sigma_{kk}^{(r)}$ for the high temperature case ($T\gg\omega_D$, where $\omega_D$ is membrane Debye frequency). In this high temperature limit, we approximate $N_{q}\approx T/q$ (we choose convenient units where the velocity of sound  $v_s=1$).  Also, we consider the energy regime $T,\omega_{D}\gg E_{b}$, and call $g_{kb}^2\xi^2\rho_{0} \rightarrow g_{kb}^2$, where $\rho_{0}$ is the (constant) vibrational density of states. 
From previous work \cite{dpc13}, we take $\xi^2\rho_{0} = 3.77 \times 10^{-5} \mbox{\r{A}}^2/\mbox{meV}$.

As a result, we conclude that the real and the imaginary parts are given as:
\begin{equation}\label{8}
{\rm Re}\Sigma_{kk}^{(r)}(E)= \frac{2g_{kb}^{2}T}{E+E_{b0}}\log\bigg|\frac{E+E_{b0}}{\epsilon}\bigg| ,   \  \  \epsilon \ll E+E_{b0}
\end{equation}
\begin{equation}\label{9}
{\rm Im}\Sigma_{kk}^{(r)}(E) = -\frac{\pi g_{kb}^{2}T}{E+E_{b0}} 
\end{equation}
where $\epsilon$ is the infrared cutoff estimated to be the minimum phonon frequency,  $\epsilon \sim v_s/L$, where $L$ is the characteristic membrane size (radius), which will be a parameter in our model.

Although the imaginary part is completely finite, Eq.~\eqref{8} shows that the real part of the finite temperature atom self-energy is log-divergent for infrared frequencies. A similar expression can be derived for the 1-loop bound state self-energy $\Sigma_{bb}$ corresponding to a Feynman diagram similar to Fig.~\ref{fig:oneloop} with $g_{kb}$ replaced with $g_{bb}$. The expressions for the real and imaginary $\Sigma_{bb}$ are given as:
\begin{equation}\label{10}
{\rm Re}\Sigma_{bb}^{(r)}(E)= \frac{2g_{bb}^{2}T}{E+E_{b0}}\log\bigg|\frac{E+E_{b0}}{\epsilon}\bigg|
\end{equation}
\begin{equation}\label{11}
{\rm Im}\Sigma_{bb}^{(r)}(E) = -\frac{\pi g_{bb}^{2}T}{E+E_{b0}} ; \ E+E_{b0} \gg \epsilon
\end{equation}

The values of $g_{bb}$ and $g_{kb}$ depend on the form of the attractive (van der Waals) potential
between the atom  and the membrane; for a H atom impinging on suspended graphene, we will take them from previous work \cite{dpc13}.
The coupling  $g_{kb}$ has a strong energy dependence, $g_{kb}=g_{kb}(E_k)$ where $E_k$ is 
the atom's  initial energy.  This energy dependence will be taken into account in our final results.
The coupling $g_{bb}^2 = 60 \mu\mbox{eV}$ is  independent of $E_k$, and in addition, is much larger than $g_{kb}^2$ (for all $E_k$ considered).

\begin{equation}
g_{bb}^2  \gg  g_{kb}^2 .
\label{newineq}
\end{equation}

The ratio of these couplings for an H atom is typically $g_{kb}^2/g_{bb}^2 \sim 10^{-2}$. 
 Because of this inequality, the bound state self-energy behavior in higher orders will be numerically much  more important than the 
 corresponding higher order contributions to  the continuum self energy. Thus we first proceed to investigate the next order in powers of $g_{bb}^2$.
 
\subsection{2-loop Bound State Self-Energy}
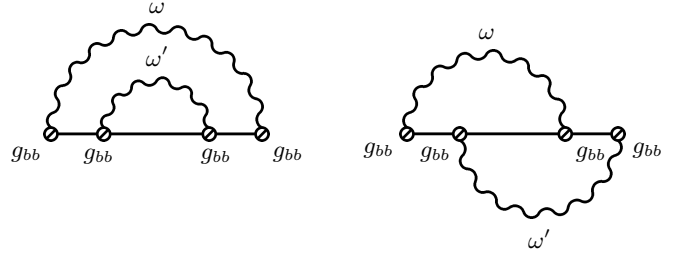
\begin{figure}[t]
\center
\begin{fmffile}{2loop}
    \begin{fmfgraph*}(100,100)
    \fmfleft{i1,d1}
    \fmfright{o1,d1}
    \fmfn{plain}{i}{2}
    \fmf{plain}{i2,v,o2}
    \fmfn{plain}{o}{2}
    \fmffreeze
    \fmf{photon,left,label=$\omega$}{i1,o1}
    \fmf{photon,left,label=$\omega'$}{i2,o2}
    \fmfblob{0.05w}{i1}
    \fmfblob{.05w}{i2}
    \fmflabel{$g_{bb}$}{i1}
    \fmflabel{$g_{bb}$}{i2}
    \fmflabel{$g_{bb}$}{o1}
    \fmflabel{$g_{bb}$}{o2}
    \fmfblob{0.05w}{o1}
    \fmfblob{.05w}{o2}
    \end{fmfgraph*}
    \hspace{1cm}
    \begin{fmfgraph*}(100,50)
    \fmfleft{i1,d1}
    \fmfright{o1,d1}
    \fmfn{plain}{i}{2}
    \fmf{plain}{i2,v,o2}
    \fmfn{plain}{o}{2}
    \fmffreeze
    \fmf{photon,left,label=$\omega$}{i1,o2}
    \fmf{photon,right,labe=$\omega'$}{i2,o1}
    \fmfblob{0.05w}{i1}
    \fmfblob{.05w}{i2}
    \fmflabel{$g_{bb}$}{i1}
    \fmflabel{$g_{bb}$}{i2}
    \fmflabel{$g_{bb}$}{o1}
    \fmflabel{$g_{bb}$}{o2}
    \fmfblob{0.05w}{o1}
    \fmfblob{.05w}{o2}
    \end{fmfgraph*}
\end{fmffile}
\vspace{1.5cm}
\caption{Two-loop bound state self-energy diagrams. Left: nested (rainbow). Right: vertex correction.}
\label{fig:twoloop}
\end{figure}

\begin{figure}
\center
\begin{fmffile}{Vertices1}
    \begin{fmfgraph*}(80,80)
    \fmfleft{i1}
    \fmfright{o1,o2}
    \fmf{photon,label=$\omega$}{i1,v2}
    \fmfblob{0.05w}{v2}
    \fmfblob{0.05w}{v1}
    \fmfblob{0.05w}{v3}
    \fmf{quark}{o1,v1,v2,v3,o2}
    \fmffreeze
    \fmf{photon,label=$\omega'$}{v1,v3}
    \end{fmfgraph*}
\end{fmffile}
%\end{center}
\caption{Vertex function $\Gamma(E,\omega)$ to be inserted appropriately in Fig.~\ref{fig:twoloop} to derive the 2-loop bound state self-energy.}
\label{fig:vertex}
\end{figure}
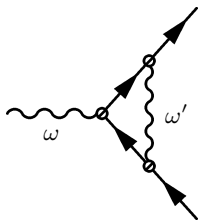

We now turn to a calculation of the 2-loop bound state self-energy corresponding to the  diagrams shown in Fig.~\ref{fig:twoloop}.
First, we evaluate the vertex correction diagram 
 by using the vertex function  $\Gamma(E,\omega)$ (see Fig.~\ref{fig:vertex}). 

Following the Feynman rules, we find the following expression for $\Gamma(E,\omega)$
\begin{equation}\label{12}
\begin{split}
\Gamma(E,\omega) &=ig_{bb}^{3}\xi^{3}\sum_{q}\int\frac{\mathrm{d}\omega'}{2\pi} \frac{T}{\omega_{q}}(-2\pi i)\bigg[\delta(\omega'-\omega_{q})\\
&\quad +\delta(\omega'+\omega_{q})\bigg]\times\bigg[\frac{1}{[E+E_{b0}-\omega'+i\eta]}\\
&\quad \times \frac{1}{[E+E_{b0}-\omega-\omega'+i\eta]}\bigg]
\end{split}
\end{equation}

%(along, up)

In the continuum limit, we call $g_{bb}^{2}\xi^{2}\rho_{0} \to g_{bb}^{2}$, so that the real and the imaginary parts of the vertex function
 $\Gamma(E,\omega)$ are written as
\begin{equation}\label{13}
{\rm Re}\Gamma(E,\omega) = \frac{2g_{bb}^{3}T}{(E+E_{b0})(E+E_{b0}-\omega)}\log\bigg|\frac{E+E_{b0}}{\epsilon}\bigg|
\end{equation}
\begin{equation}\label{14}
{\rm Im}\Gamma(E,\omega) = -\frac{\pi g_{bb}^{3}T}{(E+E_{b0})(E+E_{b0}-\omega)}
\end{equation}

Using Eqs.~(\ref{13}) and (\ref{14}), we find the contribution to the vertex correction diagram. The analytical expression for $\Sigma_{bb}$ is written as
\begin{equation}\label{15}
\Sigma_{bb}^{(2)} = ig_{bb}\xi\int\frac{\mathrm{d}\omega}{2\pi}G_{bb}(E-\omega)\Gamma(E,\omega)D^{<}(\omega)
\end{equation}
Substituting, we conclude that the real part of $\Sigma_{bb}^{(2)}$ is given by
\begin{equation}\label{16}
{\rm Re}\Sigma_{bb}^{(2)} = \frac{2g_{bb}^{4}T^{2}}{(E+E_{b0})^{3}}\bigg[\log\bigg|\frac{E+E_{b0}}{\epsilon}\bigg|\bigg]^{2}
\end{equation}

A similar expression can be easily  obtained for the contribution from the rainbow diagram. Thus, the above calculations show that  ${\rm Re}\Sigma_{bb}$ for both the 1-loop and the 2-loop is plagued by log and log-squared infrared divergences, respectively.

Therefore, at the 2-loop level,  ${\rm Re}\Sigma_{bb}(E)$ is given by
\begin{equation}\label{17}
\begin{split}
{\rm Re}\Sigma_{bb}(E) &=\frac{2g_{bb}^{2}T}{E+E_{b0}}\log\bigg|\frac{E+E_{b0}}{\epsilon}\bigg| \\
&\quad + 2 \times \frac{2g_{bb}^{4}T^{2}}{(E+E_{b0})^{3}}\bigg[\log\bigg|\frac{E+E_{b0}}{\epsilon}\bigg|\bigg]^{2} + \cdots
\end{split}
\end{equation}

Given the above structure of the expansion, it is clear that in the infrared limit $(E+E_{b0})/\epsilon \gg 1$, 
 resummation of the series must  be performed to obtain reliable results. This turns out to be possible and the infrared dynamics of the result
 is equivalent to that of the exact Green's function, $G_{bb}(E)$,  of the independent boson model \cite{mahan}. 
We can show that one can use Dyson's Equation $G_{bb}(E) = G^{(0)}_{bb}(E)/(1-G_{bb}^{(0)}(E)\Sigma_{bb}(E))$, where $G_{bb}^{(0)}(E)$ and $\Sigma_{bb}(E)$ are the unperturbed bound state Green's function and the bound state self-energy, respectively,  and then derive a perturbative expression for $G_{bb}(E)$ that matches exactly  the perturbative structure of the exact bound state Green's function corresponding to the IBM.
This is indeed natural since both the IBM and the part of our model involving $bb$ transitions describe physically equivalent situations
(a phonon bath coupled to a single particle).

%=== to be rewritten ===
%However, due to the presence of the second  ($kb$) channel, our model can be  viewed as a generalization of the IBM with two coupling constants $g_{kb}$ and $g_{bb}$.  Due to the strong inequality Eq.~\eqref{newineq} however, the infrared behavior in the $kb$ channel is much
%less important and visible in the final result. To be sure, a singular logarithmic structure appears in that channel as well, similar to
%the structure of Eq.~\eqref{17}, but with different coupling constants: at one loop the coupling is $g_{kb}^2$ (Eq.~\eqref{8}),
%while the dominant two-loop contribution, derived in Appendix A, is of order $g_{kb}^2 g_{bb}^2$ (Eq.~\eqref{twoloopappendix}).
%Thus if we compare the contributions order by order, we find that the $kb$ self-energy piece is smaller than the $bb$ piece
%by an amount $g_{kb}^2/g_{bb}^2 \sim 10^{-2}$, based on calculations for H atom. Such contributions will be neglected from now on.
%Additional evidence in favor of this overall strategy is based on analysis of higher-order contributions to the two vertices, 
%$g_{kb}^2$ and $g_{bb}^2$, as performed in Appendix B. We find that the effective vertex $g_{bb}^2$ grows,   while
%$g_{kb}^2$ decreases under renormalization (i.e. upon including higher order singular corrections). These results provide further
%justification in favor of  asymmetric treatment of the two couplings (channels) within the model.

%===== new paragraph
With the presence of the second  ($kb$) channel, our model can be  viewed as a generalization of the IBM with two coupling constants $g_{kb}$ and $g_{bb}$; however, due to the strong inequality Eq.~\eqref{newineq}, the infrared behavior originating from  higher order processes in
the $kb$ channel is strongly suppressed and will be neglected.   For example, one-loop (logarithmic) corrections to the $b$ channel propagator
due to mixing with the $k$ channel are of order $g_{kb}^2$, much smaller than the pure $bb$ channel contribution $g_{bb}^2$
calculated above. Thus the relative  contribution of these processes is smaller by a factor  of 
$g_{kb}^2/g_{bb}^2 \sim 10^{-2}$, based on calculations for H atom. Higher order processes are suppressed even stronger.

On the other hand we are ultimately interested in the Green's function of the $k$ channel (Section~\ref{sec:stick}), and its dominant  perturbative correction,
 as outlined in Appendix A, originates at order $g_{kb}^2 g_{bb}^2$. This is the dominant part in a sense that it is much
 larger than the pure mixing contribution of order $g_{kb}^4$, which can be neglected due to the same reasoning as above.
 Additional evidence in favor of this overall strategy is based on analysis of higher-order contributions to the two vertices, 
$g_{kb}^2$ and $g_{bb}^2$, as performed in Appendix B. We find that the effective vertex $g_{bb}^2$ grows,   while
$g_{kb}^2$ decreases under renormalization (i.e. upon including higher order singular corrections). These results provide further
justification in favor of  asymmetric treatment of the two couplings (channels) within the model.
 Thus we will follow the strategy of keeping the lowest necessary power of $g_{kb}^2$ while treating the $bb$ channel
 non-perturbatively.

\section{Bound state Green function within the Independent Boson Model}
\label{sec:ibm}

Based on the previous analysis we proceed to calculate the exact $bb$ Green's function which will provide
the dominant contribution to the atom damping rate, to be calculated in the next Section.
As already mentioned, if we consider only the bound state $|b\rangle$ contributions, we have the same Hamiltonian as that of the IBM:
\begin{equation}\label{18}
H = -E_{b0}b^{\dagger}b  +\sum_{q}\omega_{q}a_{q}^{\dagger}a_{q}- g_{bb}b^{\dagger}b\sum_{q}\xi_{q}(a_{q}+a_{q}^{\dagger})
\end{equation}
The exact  Green's function $G_{bb}(t)$ corresponding to Eq.~\eqref{18} is given as: \cite{mahan}
\begin{equation}\label{19}
G_{bb}(t)=-ie^{-it(-E_{b})}e^{-\phi(t)}
\end{equation}
where 
\begin{equation}\label{20}
\phi(t) = \sum_{q}\bigg(\frac{g_{bb}\xi}{\omega_{q}}\bigg)^2\bigg[N_{q}(1-e^{i\omega_{q}t}) + (N_{q}+1)(1-e^{-i\omega_{q}t})\bigg]
\end{equation}
and the binding energy $E_b$ is defined $E_b=E_{b0}+g_{bb}^2\sum_q {1\over \omega_q}$, shifted by the phonon interaction.

In the high temperature approximation, Eq.~\eqref{20} then becomes
\begin{equation}\label{22}
\phi(t) = \sum_{q}\frac{2g_{bb}^2\xi^2}{q^2}\frac{T}{q}\bigg[1-\cos(qt)\bigg]
\end{equation}
Substituting Eq.~\eqref{22} in Eq.~\eqref{19}, the Green's function  $G_{bb}(t)$ takes up the form:
\begin{equation}\label{23}
G_{bb}(t)=-ie^{-it(-E_{b})}\times \exp\bigg[-\sum_{q}\frac{2g_{bb}^2\xi^2T}{q^3}[1-\cos(qt)]\bigg]
\end{equation}
In the continuum limit, we obtain
\begin{equation}\label{24}
G_{bb}(t)=-ie^{itE_{b}} \times \exp\bigg[-2g_{bb}^{2}T{\int_{\epsilon}^{\omega_{D}}\bigg[\frac{1-\cos(q t)}{q^3}\bigg]\mathrm{d}q}\bigg]
\end{equation}

The integral in  Eq.~\eqref{24} can be expressed in terms of known functions
\begin{equation}\label{25}
\begin{split}
{\int_{\epsilon}^{\omega_{D}}\frac{[1-\cos(q t)]}{q^3}\mathrm{d}q} &= -\frac{1}{2q^{2}} + \frac{\cos(qt)}{2q^{2}}\\
&\quad + \frac{1}{2}t^{2}\text{Ci}(qt) - \frac{t\sin(qt)}{2q}\bigg|_{\epsilon}^{\omega_{D}}
\end{split}
\end{equation}
where the function $\text{Ci}(x)$ has the following expansion\cite{stegun}  for  $x \ll 1$,
\begin{equation*}
\text{Ci}(x)= \gamma + \log|x| + \sum_{n=1}^{\infty}\frac{(-1)^{n}x^{2n}}{2n(2n)!}
\end{equation*}
and  $\gamma$ is the Euler-Mascheroni constant.

The integrand of Eq.~\eqref{25} oscillates and decays rapidly with increasing $q$, so the contribution at the upper limit of integration is negligible
(especially since $\omega_D/\epsilon \gg 1$). The exponentiated function which appears in Eq.~\eqref{24}
oscillates as a function of time around the constant value  $\exp{(-g_{bb}^2T/\epsilon^2)}$ which is due to the first
term on the right-hand side of Eq.~\eqref{25}. This number is vanishingly small for all reasonable values of the cutoff and the other constants.

Next, we take the Fourier transform of Eq.~\eqref{24}, 
\begin{equation}\label{26}
\begin{split}
G_{bb}(E+E_{b})&= -i\int_{0}^{\infty} dt e^{it(E+E_{b})}\\
&\quad \times \exp\bigg[-2g_{bb}^{2} T\int_{\epsilon}^{\omega_{D}}\bigg[\frac{1-\cos(qt)}{q^3}\bigg]dq\bigg] 
\end{split}
\end{equation}
Before performing a full numerical evaluation, it is useful to estimate the decay  of the envelope of oscillations.
This can be done for large times, but subject to  the limit $t \ll 1/\epsilon$ (keeping in mind that $\epsilon$ is small), so that in the integral the largest
contribution comes from momenta $qt \ll 1$, and the logarithmic term dominates.  In this case, Eq.~(\ref{26}) reduces to
\begin{equation}\label{27}
G_{bb}(E+E_{b}) \approx -i\int_{0}^{\infty}\mathrm{d}t e^{i(E+E_{b})t}\exp\bigg[-g_{bb}^{2}Tt^{2}\log\bigg|\frac{1}{t\epsilon}\bigg|\bigg]
\end{equation}
In this regime, the oscillations are not visible.
We see that for our model, the damping factor is given as $f(t)\approx \exp\bigg[-g_{bb}^{2}Tt^{2}\log\bigg|1/(t\epsilon)\bigg|\bigg]$ which has a different structure than the case of 3D QED, where the damping factor is $f(t)\approx \exp[-\alpha Tt\log(\omega_{p}t)]$ with $\omega_{p}$ and $\alpha$ being the plasma frequency and the fine structure constant.  \cite{qed}\\

\begin{figure}
\center
\includegraphics[width=\columnwidth]{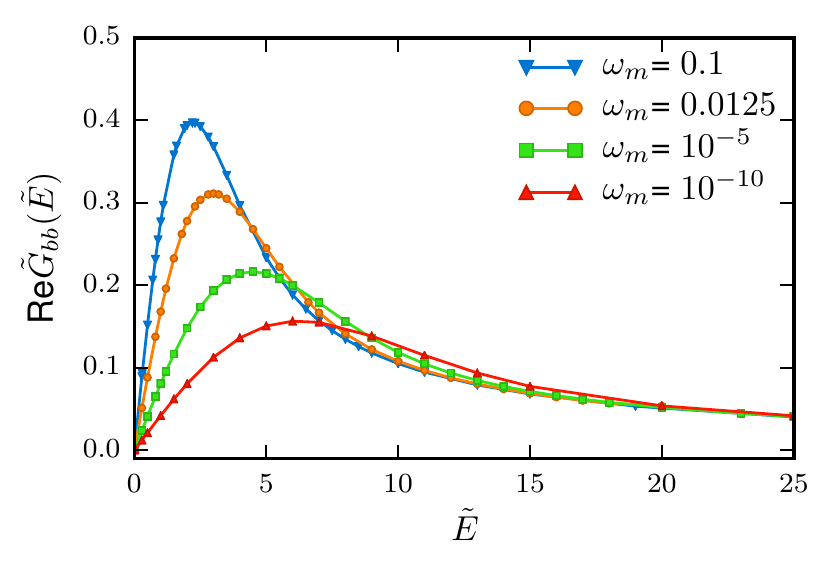}
%FIG. 6. Variation of the real part of the dimensionless Green function in the bound state , $\tilde{G_{bb}}$ for various different $\omega_{m}$.\\
\caption{Variation of the real part of the dimensionless Green's function in the bound state $\tilde{G}_{bb}$ for  different values
of the effective infrared cutoff $\omega_{m}$.}
\label{fig:ReG}
\end{figure} 

\begin{figure}
\center
\includegraphics[width=\columnwidth]{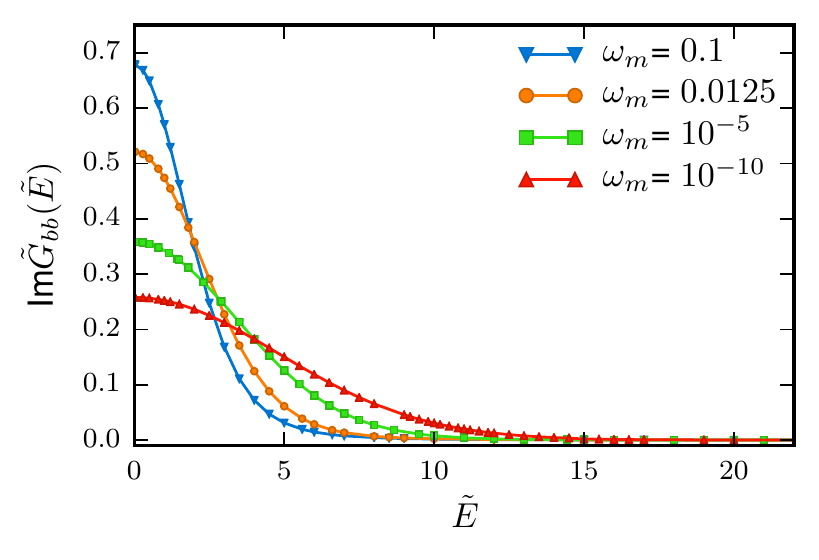}\
\caption{The imaginary part of the (dimensionless) Green function in the bound state $\tilde{G}_{bb}$ vs. energy $\tilde E$ for different
 values of the  infrared cutoff $\omega_{m}$.}
\label{fig:ImG}
\end{figure}

Next, we consider the more general case and numerically integrate Eq.~\eqref{26}. We use the following transformation of variables: $(E+E_{b})t=x$ and $q/(E+E_{b}) = y$ and under the approximation that $\omega_{D}\gg E_{b},E$, we rewrite a non-dimensional form of the bound state Green's function $\tilde{G}_{bb}$:
\begin{equation}\label{28}
\begin{split}
\tilde{G}_{bb}(\tilde{E}) &=-\frac{i}{\sqrt{\lambda}}\int_{0}^{\infty}dx \frac{e^{ix}}{\tilde{E}}\\
&\quad \times \exp\bigg[-\frac{1}{\tilde{E}^{2}}\int_{\frac{\omega_{m}}{\tilde{E}}}^{\infty}\bigg[\frac{ 1-\cos(yx)}{y^3}\bigg]dy\bigg] .
\end{split}
\end{equation}
Here we define the dimensionless cutoff $\omega_{m}$ and  energy $\tilde{E}$ in the following convenient way:
\begin{equation}
\lambda = 2g_{bb}^{2}T, \ \ \  \omega_{m} = \epsilon / \sqrt{\lambda}, \ \  \  \tilde{E} = (E+E_{b})/\sqrt{\lambda} \ .
\label{newdefs}
\end{equation}

The real and the imaginary parts of Eq.~\eqref{28} are given as follows:
\begin{equation}\label{29}
\begin{split}
{\rm Re}\tilde{G}_{bb}(\tilde{E}) &=\frac{1}{\sqrt{\lambda}}\int_{0}^{\infty}dx \frac{\sin (x)}{\tilde{E}}\\
&\quad \times \exp\bigg[-\frac{1}{\tilde{E}^{2}}\int_{\frac{\omega_{m}}{\tilde{E}}}^{\infty}\bigg[\frac{ 1-\cos(yx)}{y^3}\bigg] dy\bigg] 
\end{split}
\end{equation}
\begin{equation}\label{30}
\begin{split}
{\rm Im}\tilde{G}_{bb}(\tilde{E}) &=-\frac{1}{\sqrt{\lambda}}\int_{0}^{\infty}dx\frac{\cos (x)}{\tilde{E}}\\
&\quad \times \exp\bigg[-\frac{1}{\tilde{E}^{2}}\int_{\frac{\omega_{m}}{\tilde{E}}}^{\infty}\bigg[\frac{ 1-\cos(yx)}{y^3}\bigg] dy\bigg] 
\end{split}
\end{equation}
We choose parameters appropriate for a graphene membrane with a physisorption well $E_{b}=40$ meV, $g_{bb}^2=60$ $\mu$eV, and $\omega_{D}=65$ meV. The variation of the imaginary and the real parts of $\tilde{G}_{bb}$ with the dimensionless infrared frequency cutoff  $\omega_{m}$ is captured in Figs.~\ref{fig:ReG} and \ref{fig:ImG} respectively.

 As discussed previously,
there exists also a time-independent factor $\exp{(-g_{bb}^2T/\epsilon^2)}$ that leads to a singular $\delta(\omega)$ function
contribution with spectral weight $\exp{(-1/2\omega_m^2)}$  to the imaginary part. This contribution is not visible in Fig.~\ref{fig:ImG} since
for the cutoff values used, the additional $\delta$-function spectral weight is vanishingly small. It can however become
appreciable upon further increase of $\omega_m$ beyond $0.2$ or so, which would take us beyond the region of validity of our model. 

We conclude that both real and imaginary parts are well-behaved but still exhibit some infrared cutoff
dependence.  Most importantly, there is no quasiparticle pole and the Green's function is damped. The residual
cutoff dependence reflects the low dimensionality of the membrane flexural fluctuations.

\section{Sticking Rates}
\label{sec:stick}

We use the numerically solved $\tilde{G}_{bb}$ to derive the renomalized 1-loop atom self-energy $\Sigma_{kk}^{r}$, which in turn is used to derive the sticking rate $\Gamma$ of the cold atom on  finite temperature graphene membranes. We recall, 
the rate of transition of the cold atom from the continuum state $|k\rangle$ to the bound state $|b\rangle$ is given as:
\begin{equation}\label{31}
\Gamma = -2Z(E_{k}) Im \Sigma_{kk}(E_{k})
\end{equation} where, Z is the renormalization factor and is given as: $Z = \bigg[1-\bigg(\partial {\rm Re}\Sigma_{kk}(E_{k})/\partial E\bigg)\bigg]^{-1}$ and $E_{k}$ is the initial atom energy, respectively. \\

From Eq.~\eqref{5}, we obtain the following expression for the atom self-energy
\begin{equation}\label{32}
\begin{split}
\Sigma_{kk}^{(1)} &= g_{kb}^2 T \int_{\epsilon}^{\omega_{D}}\bigg[\frac{1}{q}\tilde{G}_{bb}\bigg(\frac{E+E_{b}-q}{\sqrt{\lambda}}\bigg)\\ 
&\quad + \frac{1}{q}\tilde{G}_{bb}\bigg(\frac{E+E_{b}+q}{\sqrt{\lambda}}\bigg)\bigg]\mathrm{d}q
\end{split}
\end{equation}

Our final results, summarized in Eqs.~\eqref{31} and \eqref{32} reflect the idea that, in order to obtain numerically accurate results,
is is sufficient to keep the lowest (first) order in the smallest coupling $g_{kb}^2$ while using the fully renormalized $G_{bb}$
which contains all orders in the strong coupling $g_{bb}^2$.

We numerically integrate Eq.~(\ref{32}) for two different infrared cut-off frequencies which physically correspond to two different sizes of the graphene membrane. We consider 1$\mu$m and 100 nm sizes. The velocity of flexural sound waves in graphene is taken to be  
 $v_{s} = \sqrt{\gamma/\sigma} =6.64 \times 10^{3}$ m/s, where $\gamma$ and $\sigma$ are defined as the out-of plane membrane tension and membrane mass density for graphene \cite{LJcutoff,dpc13,michel}, so that the physical cut-off corresponding to the two above mentioned membrane sizes are calculated as $\epsilon = 4.33 \times 10^{-3}$ meV and 0.043 meV,  respectively.

 It should be mentioned that anharmonic effects in the flexural phonon dispersion can become important as temperature increases,
 and they are a subject of current research\cite{katsnelson,amorim,zakharchenko,roldan}; however, if the tension $\gamma$ is large enough
 such effects are naturally suppressed. The tension value we use from Refs.~[\onlinecite{dpc13,lepetit-jackson}] is fairly large
 but lies in the border region where anharmonic corrections could become noticeable; detailed studies of such effects are  beyond the scope
 of the present work.

\begin{figure}
\center
\includegraphics[width=\columnwidth]{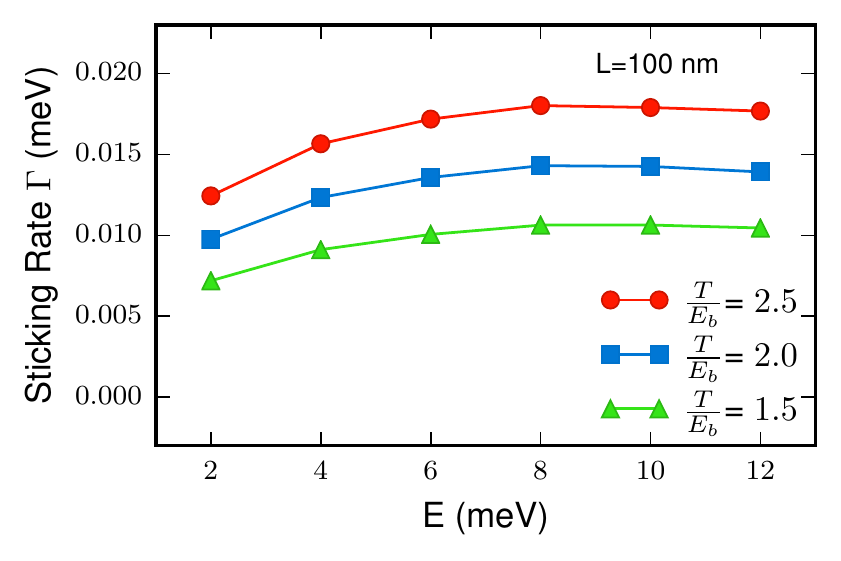}
\caption{The sticking rate $\Gamma(E)$ of the cold atom on graphene membrane  ($L=$ 100 nm),
as a function of the initial atom kinetic energy $E$. An increase in $\Gamma$ is observed for increasing temperature $T$ of the membrane.}
\label{fig:stick100}
\end{figure}

Now we present the numerical studies  for the above-mentioned membrane sizes. For each membrane size, we calculate the dependence of $\Gamma$ on temperature. For a membrane size of 100 nm, the dimensionless lower cut-off $\omega_{m}$ for the selected temperatures 1160K, 928K  and 696K  are given as 0.0125, 0.0141, 0.01628, respectively. 
In Fig.~\ref{fig:stick100} we see that the sticking rate increases with  increasing  temperature, a reflection of the physics of damping.  A higher temperature corresponds to lower physical cut-off and hence a much broader curve of Im $\tilde{G}_{bb}$ 
(see Fig.~\ref{fig:ImG}). The broadening of the curve implies more damping and hence a higher decay rate. 

\begin{figure}
\center
\includegraphics[width=\columnwidth]{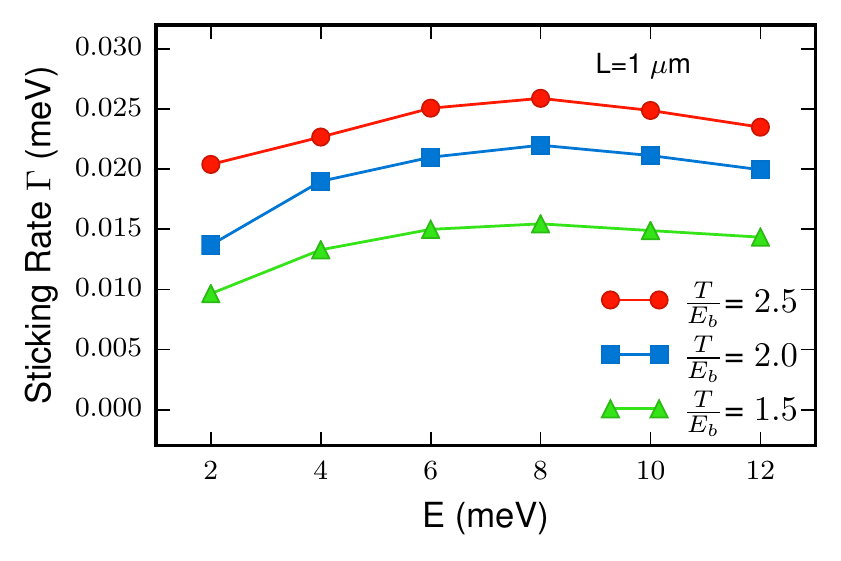}
\caption{For a membrane size of $L=1 \mu$m, $\Gamma$ is seen to increase even further with temperature $T$, which is an indication of the increasing damping of the atom wave function.}
\label{fig:stick1000}
\end{figure}

A similar trend is observed for the membrane size of 1 $\mu$m (Fig.~\ref{fig:stick1000}) where $\omega_{m}$ are given as 0.00125, 0.001397 and 0.00163 for the above-mentioned temperatures.

For comparison, the transition rate can also be estimated using Fermi's golden rule (GR)  for both in the  zero-temperature and finite temperature cases. 
The golden rule is equivalent to first-order in perturbation theory (in $g_{kb}^2$) and obviously does not contain
the additional complex physics related to infrared effects discussed previously.

For the zero temperature case
\begin{equation}\label{33}
\Gamma(T=0) = 2\pi \sum_{f}|\langle f|H_{c}|i\rangle|^{2}\delta(E_{f}-E_{i})
\end{equation}\\
Here, we use the initial state and energy as: $|i\rangle = |{k}\rangle|0\rangle$ and $E_{i} = E_{k}$.  The final state and energy is given as: $|f\rangle = |b\rangle|1_{q}\rangle$, $E_{f} = -E_{b0}+\omega_{q}$. The coupling term is given as $H_{c}= -g_{kb}(c_{k}^{\dagger}b + b^{\dagger}c_{k})\sum_{q}\xi_{q}(a_{q}+a_{q}^{\dagger}) - g_{bb}b^{\dagger}b\sum_{q}\xi_{q}(a_{q}+a_{q}^{\dagger})$.
Therefore, in the continuum limit, the sticking rate $\Gamma$ reduces to:
\begin{equation}\label{34}
\Gamma_{GR}(T=0) = 2\pi g_{kb}^2
\end{equation}\\

\begin{figure}
\includegraphics[width=\columnwidth]{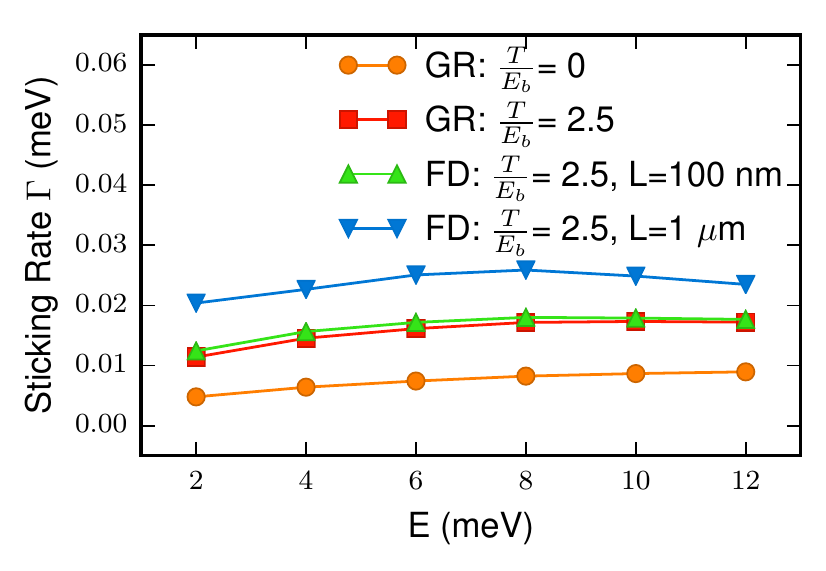}
\caption{With the increase in the cut-off $\omega_{m}$ (or decrease in the size of the membrane), the sticking rate $\Gamma$ is seen to decrease. However, we anticipate that within our model $\Gamma$ will not be smaller than the golden rule (GR) results derived for that specific temperature.}
\label{fig:stickcomp}
\end{figure}

Similarly, using the golden rule, an expression for $\Gamma$ can be derived for finite temperature. For $T\gg E_{b0}$, 
we obtain
\begin{equation}\label{35}
\Gamma_{GR}(T) = 2\pi \sum_{f}|\langle f|H_{c}|i\rangle|^{2}\delta(E_{f}-E_{i})N_{q}
\end{equation}\\
where $N_{q}$ is the equilibrium phonon number.
In the limit of $\omega_D \ll T$, we have $N_{q}\approx T/q$, and therefore:
\begin{equation}\label{36}
\Gamma_{GR}(T) = 2\pi\sum_{q}g_{kb}^2\xi^{2}\delta(-E_{b0}+\omega_{q}-E_{k})\frac{T}{q}
\end{equation}\\
In the continuum limit, we find the finite-temperature sticking rate from Fermi's golden rule as
\begin{equation}\label{37}
\Gamma_{GR}(T) = \frac{2\pi g_{kb}^2T}{(E_{k}+E_{b0})}
\end{equation}
We now compare the transition rates obtained from the golden rule both for zero and finite temperatures with the sticking rates obtained using
 $\tilde{G}_{bb}$ for the already mentioned $\omega_{m}$. We see in Fig.~\ref{fig:stickcomp} that the sticking rates derived by incorporating 
 $\tilde{G}_{bb}$
 is enhanced compared to the golden rule results, which is natural since it reflects additional damping arising from higher-order processes.

\section{Summary and Discussion}
\label{sec:conclusions}

In summary, we have considered the infrared dynamics of atoms interacting with a graphene membrane at finite temperature.
This problem exhibits particularly severe infrared divergences order by order in perturbation theory, due to the singular nature
of low-energy flexural phonon emission. Our model can be viewed as a two-channel generalization of the independent boson model,
with a much weaker atom-phonon coupling constant in one of the channels relative to the other. This allows us to take
advantage of the exact non-perturbative solution of the IBM in  the stronger channel while
treating the other one perturbatively. In the low-energy limit, the  exact solution can be viewed as a resummation  (exponentiation) of the most
divergent diagrams in the perturbative expansion, which we have checked explicitly. As a result of this procedure we obtain
the atom Green's function which we use to calculate the atom damping rate, in turn related to the  quantum sticking
rate. A characteristic feature of our results is that the Green's function retains some infrared cutoff dependence, which
is relatively weak but still detectable by relating the infrared cutoff to the inverse membrane size. We provide detailed
predictions for the sticking rate  of H atoms as a function of temperature and size. Sticking is generally enhanced relative to the conventional
Fermi golden rule result (which is equivalent to the lowest, 1-loop perturbative term) which is natural since
higher order processes are required to increase damping at finite temperature.
 Although we observe an enhancement from the golden rule result, we still see the trend of decreasing sticking rates for low energies, contrary to some recent results, where numerical calculations predict an increased sticking rate with low incident energies \cite{LJcutoff}. 
 
It is also worth noticing, as we mention in the Introduction, that our approach is very similar in spirit to the calculation of  fermion damping rates in ``hot" QED and QCD,
where the gauge structure of the theory is not particularly important as far as infrared properties are concerned.
This problem has a long history, and the relevant theoretical approach, based on the finite temperature version of the Bloch-Nordsieck
method, relies on exponentiation of infrared-divergent perturbation series (and is thus similar to the solution of the independent
boson model in solid state theory.) The role of the long-range gauge propagator is played in our case by the phonon propagator.
There are also important differences between our results and those in hot gauge theories. One difference, which
has experimental consequences for the damping rate, is the residual dependence on the infrared cutoff, which can be traced
to the quasi-1D nature of our problem (in the sense that a normally incident atom excites only axisymmetric flexural phonons). 
On the other hand, our perturbative expansion does not contain any polarization
loop corrections (which are  important ingredients of hot gauge theories), since in the case of a single atom interacting with phonon bath, those are completely absent from the theory.

We envisage applications of our approach to related physical systems, such as  graphene under additional uniaxial strain, and other atomically
thin materials, for example dichalcogenides and similar systems. In  these materials various types of strain are expected to exist \cite{Maria},
as well as larger (compared to graphene) spin-orbit  interactions. Additional uniaxial strain for example also affects
strongly the van der Waals potential near the surface \cite{Nathan}.  Therefore the atom damping rate is expected to be very sensitive
to the physical characteristics of the atom-surface interactions, such as the strain-modified  shape of the phonon flexural modes and the
van der Waals interactions between atoms and surfaces which determine the bound state energies and corresponding atom-phonon
coupling parameters.

\section{acknowledgments}

Sanghita Sengupta would like to thank Nathan S. Nichols and Adrian Del Maestro for their help with the figures in the manuscript. The research of V. N. Kotov was supported
by the U.S. Department of Energy (DOE) grant DE-FG02-08ER46512.

%\section{APPENDIX}
\appendix
\section{2-loop Atom Self-energy}
\label{sec:twoloop}
\begin{figure}[h]
\center
\begin{fmffile}{2loopAtom}
    \begin{fmfgraph*}(100,100)
    \fmfleft{i1,d1}
    \fmfright{o1,d1}
    \fmfn{plain}{i}{2}
    \fmf{plain}{i2,v,o2}
    \fmfn{plain}{o}{2}
    \fmffreeze
    \fmf{photon,left,label=$\omega$}{i1,o1}
    \fmf{photon,left,label=$\omega'$}{i2,o2}
    \fmfdot{i1}
    \fmfblob{.05w}{i2}
    \fmflabel{$g_{kb}$}{i1}
    \fmflabel{$g_{bb}$}{i2}
    \fmflabel{$g_{kb}$}{o1}
    \fmflabel{$g_{bb}$}{o2}
    \fmfdot{o1}
    \fmfblob{.05w}{o2}
    \end{fmfgraph*}
    \hspace{1cm}
    \begin{fmfgraph*}(100,50)
    \fmfleft{i1,d1}
    \fmfright{o1,d1}
    \fmfn{plain}{i}{2}
    \fmf{plain}{i2,v,o2}
    \fmfn{plain}{o}{2}
    \fmffreeze
    \fmf{photon,left,label=$\omega$}{i1,o2}
    \fmf{photon,right,labe=$\omega'$}{i2,o1}
    \fmfdot{i1}
    \fmfblob{.05w}{i2}
    \fmflabel{$g_{kb}$}{i1}
    \fmflabel{$g_{bb}$}{i2}
    \fmflabel{$g_{kb}$}{o1}
    \fmflabel{$g_{bb}$}{o2}
    \fmfdot{o1}
    \fmfblob{.05w}{o2}
    \end{fmfgraph*}
\end{fmffile}
\vspace{1.5cm}
\caption{2-loop atom self-energy diagrams: rainbow (left) and vertex correction (right).}
\label{fig:twoloopappendix}
\end{figure}
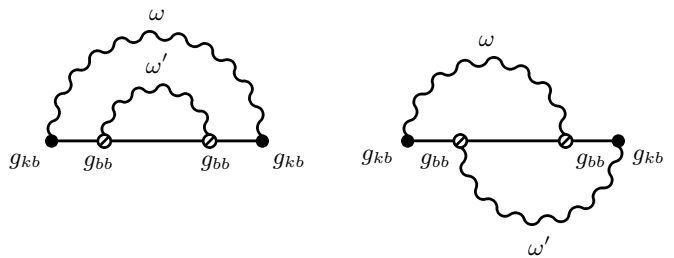

Here  we calculate the 2-loop atom self-energy corresponding to the  diagrams shown in Fig.~\ref{fig:twoloopappendix}.
These are the leading diagrams with two loops which reflect the change of the self-energy in the open $k$ channel due
to the influence of the $b$ channel.

We begin our calculation by deriving an analytical expression for the vertex function $\Gamma(E,\omega)$ 
which is given by a diagram topologically similar to Fig.~\ref{fig:vertex}, but in the $kb$ channel:
\begin{equation}
\begin{split}
\Gamma(E,\omega) &=ig_{bb}^{2}\xi^{2}g_{kb}\xi\sum_{q}\int\frac{\mathrm{d}\omega'}{2\pi} \frac{T}{\omega_{q}}(-2\pi i)\bigg[\delta(\omega'-\omega_{q})\\
&\quad +\delta(\omega'+\omega_{q})\bigg]\times \bigg[\frac{1}{[E+E_{b0}-\omega'+i\eta]}\\
&\quad \times \frac{1}{[E+E_{b0}-\omega-\omega'+i\eta]}\bigg]
\end{split}
\end{equation}
In the continuum limit, we have,
\begin{equation}
\begin{split}
\Gamma(E,\omega) &=g_{bb}^{2}\xi^{2}g_{kb}\xi\rho_{0}\int_{\epsilon}^{E+E_{b0}}\bigg[ \frac{1}{(E+E_{b}-q)}\\
&\quad \times \frac{1}{(E+E_{b0}-\omega-q+i\eta)}\\
&\quad + \frac{1}{(E+E_{b0}+q)(E+E_{b0}-\omega+q+i\eta)}\bigg]\frac{\mathrm{d}q}{q}
\end{split}
\end{equation}
Under the approximation, $q \ll E+E_{b0}$ and calling $g_{bb}^{2}\xi^{2}\rho_{0} \to g_{bb}^{2}$, the real and the imaginary parts of the vertex function $\Gamma(E,\omega)$ are written as:
\begin{equation}
{\rm Re}\Gamma(E,\omega) = \frac{2g_{bb}^{2}Tg_{kb}\xi}{(E+E_{b0})(E+E_{b0}-\omega)}\log\bigg|\frac{E+E_{b0}}{\epsilon}\bigg|
\end{equation}
Similarly, the imaginary part is given as
\begin{equation}
{\rm Im}\Gamma(E,\omega) = -\frac{\pi g_{bb}^{2}Tg_{kb}\xi}{(E+E_{b0})(E+E_{b0}-\omega)}
\end{equation}
Using the vertex function $\Gamma(E,\omega)$, we derive the contribution from the vertex-corrected self-energy first. The analytical expression 
can be written as
\begin{equation}
\Sigma_{kk}^{(2)} = ig_{kb}\xi\int\frac{\mathrm{d}\omega}{2\pi}G_{bb}(E-\omega)\Gamma(E,\omega)D^{<}(\omega)
\end{equation}
Performing the calculation, we find an expression for the real part of $\Sigma_{kk}^{(2)}$:
\begin{equation}
{\rm Re}\Sigma_{kk}^{(2)} = \frac{2g_{kb}^{2}g_{bb}^{2}T^{2}}{(E+E_{b0})^{3}}\bigg[\log\bigg|\frac{E+E_{b0}}{\epsilon}\bigg|\bigg]^{2}
\label{twoloopappendix}
\end{equation}
A similar expression is derived for the contribution from the rainbow diagram. Thus, the above calculations show that the real part of $\Sigma_{kk}$ at two-loop order is log squared infrared divergent.

\section{Vertex Renormalization}
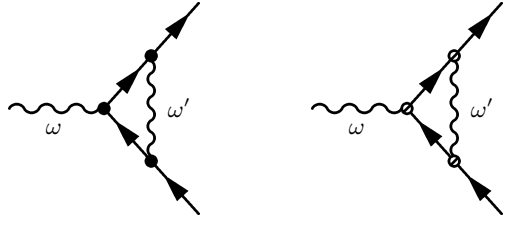
\begin{figure}[t]
\vspace{1cm}
\center
\begin{fmffile}{VR}
  \begin{fmfgraph*}(80,80)
    \fmfleft{i1}
    \fmfright{o1,o2}
    \fmf{photon,label=$\omega$}{i1,v2}
    \fmfdot{v2}
    \fmfdot{v1}
    \fmfdot{v3}
    \fmf{quark}{o1,v1,v2,v3,o2}
    \fmffreeze
    \fmf{photon,label=$\omega'$}{v1,v3}
    \end{fmfgraph*}
    \hspace{1cm}
    \begin{fmfgraph*}(80,80)
    \fmfleft{i1}
    \fmfright{o1,o2}
    \fmf{photon,label=$\omega$}{i1,v2}
    \fmfblob{0.05w}{v2}
    \fmfblob{0.05w}{v1}
    \fmfblob{0.05w}{v3}
    \fmf{quark}{o1,v1,v2,v3,o2}
    \fmffreeze
    \fmf{photon,label=$\omega'$}{v1,v3}
    \end{fmfgraph*}
\end{fmffile}
%\vspace{5mm}
%
%(a)
%\hspace{4cm}
%(b)\\
\caption{Vertex diagrams corresponding to transitions from:  $|k\rangle \rightarrow |b\rangle$ states  (left), and   $|b\rangle \rightarrow |b\rangle$ states (right).}
\label{fig:twovertices}
\end{figure}

Now we  calculate the vertex renormalization for the two  different types of vertices in our model. The one loop vertex diagrams are shown in Fig.~\ref{fig:twovertices}.\\

\begin{figure}[h]
\center
\begin{fmffile}{CD}
    \begin{fmfgraph*}(100,100)
    \fmfleft{i1} 
    \fmfright{o1,o2}
    \fmfdot{v2}
    \fmfblob{0.05w}{v1}
    \fmfblob{0.05w}{v4}
    \fmfdot{v5}
    \fmf{photon}{i1,v3}
    \fmf{quark}{o1,v1,v2,v3,v4,v5,o2}
    \fmffreeze
    \fmf{photon}{v1,v4}
    \fmf{photon,rubout}{v2,v5}
    \end{fmfgraph*}
\end{fmffile}
\caption{Higher order crossed vertex corrections to the $kb$ vertex.}
\label{fig:crossedvertex}
\end{figure}
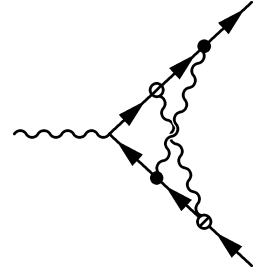

The corrections to the vertices are represented as $(g_{kb} + \delta \Gamma^{(1)}_{kb})$ and $(g_{bb}+ \delta \Gamma^{(1)}_{kb})$.
By evaluating the corresponding diagrams we obtain for the infrared-divergent parts:
\begin{equation}
\delta \Gamma^{(1)}_{kb}(E) = - \frac{g_{kb}^{3}T}{\pi(E+E_{b0})^{2}}\log\bigg|\frac{E+E_{b0}}{\epsilon}\bigg|,
\end{equation}
\begin{equation}
\delta \Gamma^{(1)}_{bb}(E) =  \frac{g_{bb}^{3}T}{\pi(E+E_{b0})^{2}}\log\bigg|\frac{E+E_{b0}}{\epsilon}\bigg|.
\end{equation}
Here, the external phonon frequency is set to zero (infrared limit), while $E$ is the external atom energy.
The most important feature of these results is that the corrections have different signs, i.e. while the $kb$ vertex decreases, the $bb$ vertex increases.

It is possible to write down and solve the corresponding Dyson equations for the fully renormalized vertex functions, which is equivalent to summing an infinite series of ladder
diagrams as is conventionally done in QED \cite{landaulifshitz4}. This results in the following expressions for the effective
vertices in the two channels:
\begin{equation}
\Gamma_{kb}(E) = \frac{g_{kb}}{1+\frac{g_{kb}^{2}T}{\pi(E+E_{b0})^{2}}\log\bigg|\frac{E+E_{b0}}{\epsilon}\bigg|},
\end{equation}
and
\begin{equation}
\Gamma_{bb}(E) = \frac{g_{bb}}{1-\frac{g_{bb}^{2}T}{\pi(E+E_{b0})^{2}}\log\bigg|\frac{E+E_{b0}}{\epsilon}\bigg|}.
\end{equation}
Again, it is clear that $\Gamma_{kb}(E) $ decreases while $\Gamma_{bb}(E)$ increases in the infrared limit. $\Gamma_{bb}(E)$ in fact contains a Landau pole, 
although due to the smallness of the effective coupling, the system never reaches the pole for  physical values of the parameters
(coupling, temperature and cutoff).

Finally, we consider  even higher order  renormalization effects. For the $kb$ vertex, the next level of complexity is represented
by the crossed vertex corrections shown in Fig.~\ref{fig:crossedvertex}. By evaluating the diagram we obtain
\begin{equation}
\delta \Gamma^{(2)}_{kb}(E) = - \frac{2g_{kb}^{3}g_{bb}^2T^2}{\pi^2(E+E_{b0})^{4}}\log^2\bigg|\frac{E+E_{b0}}{\epsilon}\bigg|,
\end{equation}
which confirms that the $kb$ vertex keeps decreasing. Corresponding results can be derived for the $bb$ vertex (which experiences an increase).
These results are conceptually important because they reaffirm the different tendencies in the two channels, although
numerically  these diagrams  are very small for physical parameter values.

%\bibliographystyle{apsrev4-1}
%\bibliography{qs}

%merlin.mbs apsrev4-1.bst 2010-07-25 4.21a (PWD, AO, DPC) hacked
%Control: key (0)
%Control: author (72) initials jnrlst
%Control: editor formatted (1) identically to author
%Control: production of article title (-1) disabled
%Control: page (0) single
%Control: year (1) truncated
%Control: production of eprint (0) enabled
%

\end{document}